\documentclass[aps,prl,groupedaddress,showkeys,doublespace]{revtex4}
\usepackage{graphicx}
\usepackage{doublespace}

\begin{document}

\title{Simple solvation potential for coarse-grained models of proteins}

\author{A. Bhattacharyay\footnote[1]{arijit@pd.infn.it}, A. Trovato\footnote[2]{antonio.trovato@pd.infn.it}, F. Seno\footnote[3]{flavio.seno@pd.infn.it}}
 
\affiliation{Dipartimento di Fisica ``G.Galilei'', Universit\`a degli Studi 
di Padova, via F. Marzolo 8, 35131 Padova, Italy}
 
\date{\today}

\begin{abstract}

We formulate a simple solvation potential based on a coarsed-grain
representation of amino acids with two spheres modeling the $C_\alpha$
atom and an effective side-chain centroid. The potential relies on a
new method for estimating the buried area of residues, based on
counting the effective number of burying neighbours in a suitable
way. This latter quantity shows a good correlation with
the buried area of residues computed from all atom crystallographic
structures. We check the discriminatory power of the solvation
potential alone to identify the native fold of a protein from a set of
decoys and show the potential to be considerably selective.

\end{abstract}

%\pacs{}


\maketitle

\newpage
\section{introduction}
The prediction of three dimensional structure of the native state of
proteins from the knowledge of their sequence of amino-acids can only
be achieved if the interaction potentials among various parts of the
peptide chain in the presence of solvent molecules are known to some
extent. In fact, approaches to protein structure prediction are based
on the thermodynamic hypothesis that the native state of a protein is
the state of lowest free energy under physiological conditions
[1]. Thus, the computation of the energy of a sequence in a given
conformation would be the fundamental step toward the solution of the
protein folding problem [2].  A rigorous approach [3] from ``first
principles'' taking into account the quantum mechanics of a huge
number of atoms constituting the protein is not practical and beyond
actual computational capabilities. An usual way [4] to avoid dealing
with too many microscopic degrees of freedom is to introduce a reduced
representation of proteins in which each amino acid is represented by
one or a few interaction sites. The main difficulty with such
simplified representations is the need of an effective energy that
captures, at least, the essential qualitative physical and chemical
features of the folding process. This involves choosing the right
contributions to the energy function as well as determining the
related potentials [see for example 2,26-30].

Since the classical work of Kauzmann [5] it is evident [6] that the
hydrophobic effect is one of the leading forces in the folding
process. Hydrophobicity arises because the side-chain of some amino
acids is not able to form hydrogen bonds with the surrounding water,
and as a consequence its solubility is low. Gordon et al. [7] have
reported a free energy difference associated with exposing such groups
to water which is comparable with the strength of hydrogen
bonds. Moreover, several experiments [8,9,10,11] have shown that non
specific interaction and placement of hydrophobic residues is a
critical determinant of protein structure.

Starting from the seminal work of Eisenberg and McLachlan [12] several
efforts have been produced to model the hydrophobic effects [13-17]
but a precise and computationally efficient method is still lacking.
The purpose of this work is to cover this gap by developing a
solvation energy for a coarse-grained model of a protein which is more
accurate than the other existing ones but still is easy for numerical
implementation in folding simulations. The energy function we propose
is the sum over all the amino-acids of a contribution given by a
factor proportional to the effective number of particles which are
screening the amino-acid from its aqueous surroundings multiplied by
an amino acid specific parameter measuring its hydrophobicity
degree. These parameters are calculated through a statistical analysis
on a training set of proteins.

An important test to validate any energy function is to check its
ability to recognize native structures among a large number of well
constructed decoys [18-21]. A number of standard decoy sets,
consisting of the native structure plus a large ensemble of simulated
protein-like structures have been established for benchmarking
purposes [22]. The performance of our method on different test sets is
better than other solvation energies [16,17] and is comparable to
scoring functions which implicitly include other energetic
contributions such as hydrogen bond, torsion angle and van der Waals
potentials [23,24].

% Furthermore the use of our solvation energy in the
%recent proposed algorithm Victor/FRST[25] allows a very high prediction
%accuracy in all the test sets.

\section{Methods}
In our model each residue is simplified as follows. The $C^\alpha$
atoms are represented by a sphere of radius 1.9 \AA. The specific
characteristic of each amino acid is captured by its side-chain which
we again model as a sphere. The center of this sphere is located at
the geometric center, ${\bf X_c}$, of all side chain atoms including
the $C_\beta$ atom, namely at:

$$
{\bf X}_c=\frac{\sum_{l=1}^{m} {\bf X}_l}{m}.
$$ 

\noindent where ${\bf X_l}$ are the positions of all the atoms in the
given side chain and $m$ is the total number of atoms in the side
chain. The radius $r_c$ of the side-chain sphere is defined as the
root mean square deviation of the position of the side-chain atoms of
that residue from ${\bf X_c}$. Relative changes in positions of side
chain atoms of a particular residue in various proteins are reflected
in the corresponding change in position of the side-chain center and
in some fluctuation of the side change radius. Since modeling side
chains as spheres is a gross simplification of side chain geometry,
the slight overlap of different sphere pairs can occur. Note that
throughout this paper we do not enforce any steric constraint.

Our main purpose is to find a way within this representation to
compute the effective number $N_i^a$ of spheres (both $C^{\alpha}$ and
side chains) which are burying the side chain of the amino-acid of
type $a$, located at position $i$ along the chain, from the
solvent. This number will then be used as an estimation of the buried
area of residue $i$ in order to define a solvation energy in this
coarse-grained perspective. Let us call $D_{jk}$ the distance between
two generic spheres $j$ and $k$ (note the difference between indexes
$j,k$ going through both $C^{\alpha}$ and side chains and residue
index $i$ going only through side chain spheres). All other spheres in
the chain, either $C^{\alpha}$ or side chains, contribute to $N_i^a$
through the relation:

\begin{equation}
N_i^a  =\mathop{\sum_{j \notin i}}_{j \notin i \pm 1}
B_{ij} S_{ij}
\label{eq1}
\end{equation}

\noindent where the sum is running only through spheres belonging to
residues not adjacent to $i$. $B_{ij}$ represents the bare
contribution by sphere $j$ to the protection of side chain $i$ from
water and it is equal to $1$ if the distance $D_{ij}$ is smaller than
$R_{ij}= r_i+r_j+\frac{r_{\rm water}}{2}$ ($r_{\rm water}=1.4\ \AA$ is
the radius of a water molecule) or $\frac{R_{ij}}{D_{ij}}$
otherwise. The threshold $R_{ij}$ has been chosen to discriminate
between two regimes. If $D_{ij}<R_{ij}$, a portion of the surface area
of side chain $i$ is screened from contact with water molecules, since
the latter can only marginally accomodate between side chain $i$ and
sphere $j$ (see Fig. 2). If $D_{ij}>R_{ij}$, this constraint
disappears but in order to take into account fluctuations, due to the
varying geometry of side chains and to the simple assumption of
spherical geometry, we assume the screening effect to be still present
and decaying as $\frac{1}{d}$.

\noindent $S_{ij}$ instead takes into account the fact that the
screening effect of sphere $j$ on side chain $i$ can be already
accounted for by the presence of other spheres of the chain.
We define $S_{ij}$ as

$$
S_{ij}= \mathop{\prod_{k\neq j}}_{k \notin i,i\pm 1} S_{ij}^k
$$

\noindent where we multiply the individual contributions $S_{ij}^k$ of
different spheres to the overlap of screening effects, given by

$$ 
S_{ij}^k=
\left\{ \begin{array}{ll}
 \left(1-M_{ijk} \right)\frac{r_j}{r_k} &   \ \ \ \ \      {\rm if} M_{ijk} >
 0.7 \\
1 &  \ \ \ \ \  {\rm if} M_{ijk} \leq 0.7 \  {\rm or}\  D_{ij} \leq D_{ik}
\end{array}
\right\}
$$

\noindent The coefficients  $M_{ijk}$ are defined as

$$
M_{ijk}= \frac{D_{ij}}{D_{ik}+D_{kj}}
$$

\noindent The $M_{ijk}=1$ limit corresponds to a complete overlap of
the screening effects (Fig.2c) so that in this case $S_{ij}^k=0$;
sphere $j$ is not contributing effectively in screening side chain $i$
from water molecules more than sphere $k$ is already doing. It is easy
to demonstrate that when three spheres of equal size are in contact
with each other, $M_{ijk}$ is equal to $0.5$ (e.g. see Fig.2a). In
such a situation no sphere is actually obstructing the other two to
come in complete contact and there is no overlap of the screening
effects. Starting from such a limiting geometry, when sphere $k^{\rm
th}$ is inserting between the other two (Fig.2b) the value of
$M_{ijk}$ is increasing. The choice of $0.7$ as the lower limit to
measure the screening overlap has been made based upon an optimization
procedure. The factor $\frac{r_j}{r_k}$ takes into account the fact
that the bigger $r_k$ the more overlapping the screening effects of
$j$ and $k$, and therefore the smaller $S_{ij}^k$. It is interesting
to note that statistics of side chain sizes, based on our
coarse-grained representation, reveals that the ratio $r_j/r_k$ is
generally less than $3$. So, $1\ge M_{ijk}>0.7$ never let
contributions to $S_{ij}^k$ be bigger than unity.

To test our method we use the Top500H database [31] of non-redundant
protein structures. It is a hand-curated set of 500 high-resolution
structures all solved by X-ray crystallography to $1.8 \ \AA$ or
better resolution. From this database, only those $220$ proteins which
do not have any discontinuity in the chain have actually been
selected. The analysis of this proteins shows that the effective
number of contacts is ranging from $0$ to about $12$ (see Fig. 3). We
took $10$ to be the upper limit in all further calculation since it
was already shown to be roughly the number of neighbours with which an
amino-acid becomes buried (see, e.g. [17,32]). Our way of taking into
account the overlap of different screening effects could indeed
produce some overcounting and whenever the counting of $N_i^a$ is
above $10$, it is put equal to it. To check the correspondence between
$N_i^a$ and buried area, we have calculated the buried area of
residues of these proteins by using on-line available program
{\it'GETAREA 1.1'} [34] (URL:
http://www.scsb.utmb.edu/cgi-bin/get\_a\_form.tcl). The number $B_i^a$
that we use from this program is the percentage of buried side-chain
surface area with respect to the ``random coil'' value per
residue. The ``random coil'' value for residue is the average
solvent-accessible surface area of $a$ in the tripeptide $Gly-a-Gly$
in an ensemble of 30 random conformations. $B_i^a$ varies from $ 0 \%$
(completely exposed) to $100 \%$ (fully buried) and it is plotted in
Fig.3 against $N_i^a$ for all amino-acids of a set of proteins (see
Results for discussions). The average buried area $\overline{B}^a$ of
each residue kind is displayed in Table A.

In order to develop the solvation energy we consider the amino-acids
to be hydrophobic if their average buried area is more than $60 \%$,
otherwise we take them to be polar. The solvation energy parameters
$E_a$ for different residue types, listed in Table A, are obtained
through the relation:

$$ E_a=(60-\bar B^a)/(\bar B_{max}-60) $$

\noindent where $\bar B_{max}= max \{\bar{B^a}\}= 86.90 \%$. In our
coarse-grained representation the overall solvation energy is then
obtained summing contribution from all residues, negative for
hydrophobic ones ($E_a<0$), and positive otherwise, which are
proportional to the residue buried area $N_i^a$ evaluated from
Eq. (\ref{eq1}):

\begin{equation}
E_{solv}=\sum_i E_a\times N_i^a,
\label{eq2}
\end{equation}

%and that of a polar residue is given as
%$$
%E_{solv}=E_a\times (10-N_i^a).
%$$

\noindent Note that no solvation energy is associated to $GLY$ in our
model, but it takes part into the calculation of solvation energy of
other residues by virtue of its $C^\alpha$ sphere.

\par Five different performance measures have been applied to assess
the ability of our solvation potential in discriminating the native
structure from native like ones. The $Rank1$ measure is the one which
is more direct in the sense it tells us about the position of the
native state among the decoys in an arrangement of increasing order of
their energy. So, $Rank1$ being equal to $1$ means the native fold is
the lowest energy conformation among the decoys. We also employ the
$Z$-score which is defined as 

$$
Z-{\rm score}=\frac{E_{native}-\bar{E}}{\sigma}
$$

where, $E_{native}$ is the energy of the native fold and $\bar{E}$ is
the average energy of decoys with a standard deviation $\sigma$. We
also employ additional performance measures to understand the
discriminatory capacity of our potential for native like structures
rather than oly for the native fold. To assess such properties we use
$log PB1$, $log PB10$ and $F.E.$ (fraction enrichment) [33] measures
to see how the present potential fares. $logPB1$ is the log
probability of selecting the best scoring conformation and is given by

$$
logPB1=log_{10}\left(\frac{R_i}{p}\right)
$$

where $R_i$ is the $RMSD$ rank of the best scoring conformation in $p$
decoys.  $logPB10$ is the probability of selecting the lowest $RMSD$
conformation among the 10 best scoring conformations
$(R_i=min\{R_1,...,R_{10}\})$. The $F.E.$ or fraction enrichment is
the percentage of the top 10\% lowest $RMSD$ conformations in the top
10\% best scoring conformations.

\section{results}

Our first test aims to verify the existence of a correlation between
the method, presented in the previous section, of counting effective
neighbours ($N_i^a$) of an aminoacid and its buried area.  To perform
this test, we have used the Top500H database [31] of non-redundant
protein structures selecting only those proteins which do not have any
discontinuity in the chain. Fig.3 shows a plot of the calculated
buried area, obtained using {\it GETAREA 1.1}, of different residues
in their native conformations versus their $N_i^a$.  The correlation
turns out to be quite good (coefficient of correlation = 0.89).

The crucial test for our method consists, however, in assessing the
ability of the solvation energy based on Eq. (\ref{eq2}) to
discriminate between native states and not native decoy conformations
for the same sequence. A good decoy set must include a large number of
conformations, some near native and other that are native-like in all
respects except the overall folded conformation and this set should be
generated independently from the evaluated scoring mechanisms to avoid
bias toward any particular selection methodology. Our solvation
potential has been tested on four such standard decoy sets:
$4state\_reduced$, $lattice\_ssfit$, $lmds$ sets from the Decoys 'R'
Us web sites (http//:dd.stanford.edu and
http//:ddcompbio.washington.edu) and the $Rosetta$ decoys from the
Baker laboratory site (http//:depts.washington.edu/bakerpg).

In Table B we arrange the average scores for the different sets
whereas the detailed account of the individual scores for every target
in various decoy sets are given in Tables C-F.  This detailed scores
are more revealing than the average ones due to the fact that some
targets which are scoring particularly bad due to some specific
reasons can affect the average score.

\par For the $4state\_reduced$, we observe that $Rank1=1$ four times
out of seven( $4/7$.) It is $2$ for the protein $2cro$ whereas $5$ and
$4$ for the proteins $4pti$ and $4rxn$, respectively.  Protein $4pti$
has three disulphide bonds which might be in conflict with the optimum
arrangement on the basis of solvation energy alone. This performance
can be compared with other two solvation potentials: the one called
{\it Chebyshev-expanded hydrophobic} potential ($CHP$) introduced by
Fein, Xia and Levitt [17] and the one introduced by Jones [16]
($SOLV$) used by Tosatto [25] to formulate a combined potential
function called $FRST$. The discriminatory predictions of $CHP$,
$SOLV$ and $FRST$ as well (although is not a solvation potential) are
reported in table G and H for the available sets and with the
available information. For the $4state\_reduced$ set, both $CHP$ and
$SOLV$ get $ Rank=1$ only in one case and both methods
with a worse average value for the parameter $Z$-score.

For the $lattice\_ssfit$ set of decoys we obtain $Rank1=1$ in 6 cases
out of 8. Again the performance is much better of $SOLV$ (there are
not available data for $CHP$) which has $Rank1=1$ in 4 cases out of 8,
and worse values for the $Z$-score, $logPB1,\ logPB10$ and $F.E.$
parameters.  In this set, the targets for which our method gives
$Rank1 \neq 1$ are the proteins $1dkt-A$ and $1trl-A$: these two
proteins are a part of a whole chain. The missing parts might have
residues which contribute to the count of $N_i^a$. Their absence can
drastically reduce the efficiency of the methods.

For the set of decoy $lmds$, $Rank1=1$ in $4$ proteins out of $8$. The
proteins {\it 1b0n-B} and $1fc2$ are the ones which show almost no
selection on the basis of this solvation potential because they are
short chains of hetero-dimers. Other two, $2ovo$ and $4pti$, have
three disulfide bonds. Interestingly, $4pti$ has a $Rank1$ measure of
the same order also in the decoy sets $4state\_reduced$ and $Rosetta$.
Again the performance is much better than for the other two solvation
energy potentials.

For the set $Rosetta$, $Rank1=1$ in 17 cases out of 23.  In this
collection of decoys, there are proteins which have residues clipped
from the chain end and thus decoys are considerably shorter than the
actual protein. We have selected those targets from $Rosetta$ which
have at most up to $8$ residues clipped from the end of a chain (in Table F
the numbers within bracket just after the target name are the numbers of 
residues clipped). We
are not considering those having more residues clipped since it can
affect proper calculation of solvation energy. It is important to note
that the targets for which the $Rank1$ measure is not very good are
mainly those which have a large number of residues clipped, even
though some of the latter have $Rank1=1$. We can not compare our
method on this decoys set with $CHP$ and $SOLV$ since these data are
not available. Nevertheless, we make a comparison with an all
atom-atom contact scoring energy proposed by McConey et al. [23] which
has a record for $Rank=1$ of 19 out of 23 with a $Z-{\rm score}=-3.6$,
performances which are just slightly better of ours. Keeping in mind
that the approach in [23] is taking care in detail of various
interactions with more than 28000 parameter it turns out that our
solvation energy has a very high degree of confidence in
discriminating native state with respect to other methods.

For all the set in which is possible to make a comparison the full
algorithm $FRST$ is performing better than our method, although
sometimes the values of the less selective parameters $logPB1,
logPB10$, and $F.E$ are comparable. This is a consequence of the fact
that $FRST$ is an algorithm which combine four different knowledge
based potentials. In addition to the solvation potential ($SOLV$),
there are the pairwise, the hydrogen bond and the torsion angle
potentials. The results we presente here suggest that, regarding the
development of refined methods for structure recognition, it should be
convenient to include our method rather than $SOLV$ in the $FRST$
algorithm.

\section{Discussion}

In the present work we have proposed a simple and efficient method to
estimate the buried area of a side chain group by counting the
effective number of atoms which are screening it from water. This
evaluation builds on the assumption of spherical symmetry of the
interacting side chain groups. It corrects for this oversimplification
by allowing the screening effect to be considered beyond the range
dictated by sphere sizes and by considering the three-body
contribution to its cooperativity. The resulting quantity is well
related to the buried area and it can be used to compute a solvation
energy which is based just on 20 energy solvation parameters, one for
each kind of amino-acid. This solvation energy works pretty well,
better than any other solvation energy, in recognizing native
structures among set of well constructed, alternative decoys. The use
of a small number of parameters makes our solvation potential more
physically transparent than other elaborate knowledge based energy
functions with a much larger parameter space.

The targets which are parts of larger chain, or have a good number of
residues clipped from the chain end or having many disulfide bonds etc
are generally those for which the present solvation potential fails to
identify the native fold.

Our potential does not implicitly include any other interaction than
the solvation effect, so it should be easily improved by combining it
with other interaction terms. Moreover, since the model we use to
represent amino-acids is quite simplified, the approach can be easily
implemented in ab-inito simulation of protein folding.

\section{Acknowledgments} We thank A. Maritan and S. Tosatto for
sitmulating discussion. This work was supported by PRIN 2005
prot. 2005027330 and the Program ``Progetti di Ateneo'' of Padua
University.

%\newpage 

{\bf Table A: Data showing average measure of buried area for Amino acids in native state's database}

\begin{center}
\begin{tabular*}{0.50\textwidth}{@{\extracolsep{\fill}}c|cccc}
Amino acid & $\overline{B}^a$ & $E_{a}$ \\
\hline
ALA & 71.55 & -0.39 \\
ARG & 59.57 & 0.01 \\
ASN & 58.77 & 0.04 \\
ASP & 55.87 & 0.14 \\
CYS & 89.80 & -1.00 \\
GLN & 57.46 & 0.09 \\
GLU & 50.44 & 0.32 \\
HIS & 70.12 & -0.34 \\
ILE & 86.90 & -0.90 \\
LEU & 85.48 & -0.86 \\
LYS & 44.30 & 0.53 \\
MET & 81.92 & -0.74 \\
PHE & 86.82 & -0.90 \\
PRO & 57.34 & 0.09 \\
SER & 61.94 & -0.07 \\
THR & 65.91 & -0.20 \\
TRP & 84.95 & -0.84 \\
TYR & 80.36 & -0.68 \\
VAL & 85.36 & -0.85 \\
\hline
\end{tabular*}
\end{center}                  
   
\newpage
{\bf Table B: Average performance measures}

\begin{center}
\begin{tabular*}{0.50\textwidth}{@{\extracolsep{\fill}}l|ccccc}
Decoy Set & Rank 1 & $Z$-score & logPB1 & logPB10 & F.E. \\
\hline
4state\_reduced & 4/7 & -2.89 & -1.19 & -1.88 & 0.28 \\
lmds & 4/8 & -2.75 & -0.48 & -1.36 & 0.14 \\
lattice\_ssfit & 6/8 & -4.06 & -0.41 & -1.55 & 0.12 \\
rosetta & 17/23 & -3.44 & -0.54 & -1.65 & 0.16 \\
\hline
Total/average & 31/46 & -3.28 & -0.85 & -1.61 & 0.17 \\
\hline
\end{tabular*}
\end{center}

{\bf Table C: Performance Measure for 4state\_reduced}

\begin{center}
\begin{tabular*}{0.50\textwidth}{@{\extracolsep{\fill}}l|ccccc}
Target & Rank 1 & $Z$-score & logPB1 & logPB10 & F.E. \\
\hline
1ctf & 1 & -3.53 & -2.10 & -2.10 & 0.55 \\
1r69 & 1 & -3.68 & -1.93 & -2.53 & 0.27 \\
1sn3 & 1 & -2.52 & -0.27 & -1.09 & 0.10 \\ 
2cro & 2 & -3.01 & -1.26 & -2.53 & 0.32 \\
3icb & 1 & -2.26 & -2.12 & -2.12 & 0.46 \\
4pti & 5 & -2.51 & -0.50 & -1.48 & 0.06 \\
4rxn & 4 & -2.71 & -0.15 & -1.88 & 0.20 \\
\hline
Total/average & 4/7 & -2.89 & -1.19 & -1.88 & 0.28 \\
\hline
\end{tabular*}
\end{center}

\newpage

{\bf Table D: Performance Measure for lmds (* excluded from the
calculation of averages. We have not considered the score of the
targets $1b0n\-B$ and $1fc2$ while calculating the averages.  )}

\begin{center}
\begin{tabular*}{0.50\textwidth}{@{\extracolsep{\fill}}l|ccccc}
Target & Rank 1 & $Z$-score & logPB1 & logPB10 & F.E. \\
\hline
1b0n-B* & 439 & 1.18 & -0.01 & -0.32 & 0.04 \\
1ctf & 1 & -3.42 & -0.36 & -0.77 & 0.08 \\
1fc2* & 409 & 0.91 & -0.31 & -2.22 & 0.07 \\
1igd & 1 & -2.87 & -0.25 & -1.39 & 0.14 \\
1shf-A & 1 & -2.90 & -0.14 & -1.09 & 0.09 \\
2cro & 1 & -3.42 & -0.60 & -1.37 & 0.18 \\ 
2ovo & 16 & -1.67 & -1.00 & -1.46 & 0.14\\
4pti & 6 & -2.24 & -0.53 & -2.06 & 0.20 \\
\hline
Total/average & 4/8 & -2.75 & -0.48 & -1.36 & 0.14 \\
\hline
\end{tabular*}
\end{center}

{\bf Table E: Performance Measure for lattice\_ssfit}

\begin{center}
\begin{tabular*}{0.50\textwidth}{@{\extracolsep{\fill}}l|ccccc}
Target & Rank 1 & $Z$-score & logPB1 & logPB10 & F.E. \\
\hline
1ctf & 1 & -5.04 & -0.06 & -0.68 & 0.14 \\
1beo & 1 & -3.67 & -0.36 & -1.13 & 0.12 \\
1dkt-A & 8 & -2.73 & -0.20 & -2.35 & 0.11 \\
1fca & 1 & -7.38 & -2.45 & -2.45 & 0.09 \\
1nkl & 1 & -4.54 & -0.01 & -2.52 & 0.08 \\
1pgb & 1 & -4.01 & -0.10 & -0.51 & 0.10 \\
1trl-A & 101 & -1.61 & -0.12 & -0.98 & 0.18 \\
4icb & 1 & -3.50 & -0.01 & -1.75 & 0.12 \\
\hline
Total/average & 6/8 & -4.06 & -0.41 & -1.55 & 0.12 \\
\hline
\end{tabular*}
\end{center}

\newpage

{\bf Table F: Performance Measure for rosetta (* excluded from the 
calculation of averages)}

\begin{center}
\begin{tabular*}{0.50\textwidth}{@{\extracolsep{\fill}}l|ccccc}
Target & Rank 1 & $Z$-score & logPB1 & logPB10 & F.E. \\
\hline
1aa2(3) & 1 & -4.18 & -0.65 & -1.70 & 0.15 \\
1acf(2) & 1 & -5.34 & -0.31 & -2.15 & 0.23 \\
1bdo(5) & 1 & -3.81 & -0.03 & -1.55 & 0.14 \\
1cc5(7)* & 407 & -0.30 & -0.12 & -1.06 & 0.12 \\
1csp(3) & 1 & -2.73 & -0.18 & -2.52 & 0.20 \\
1ctf(1) & 1 & -3.49 & -0.06 & -1.34 & 0.14 \\
1eca(4) & 1 & -3.67 & -0.21 & -2.00 & 0.13 \\
1erv(0) & 1 & -4.20 & -0.56 & -3.00 & 0.21 \\
1kte(5) & 1 & -2.67 & -1.55 & -1.55 & 0.12 \\
1lfb(8) & 40 & -1.47 & -0.97 & -1.68 & 0.12 \\
1mbd(6) & 1 & -4.37 & -1.25 & -1.44 & 0.14\\
1msi(5) & 1 & -4.61 & -1.10 & -2.00 & 0.15\\
1pal(8) & 4 & -2.07 & -0.25 & -1.92 & 0.14\\
1pdo(8) & 1 & -3.79 & -0.90 & -1.33 & 0.20\\ 
1ptq(7)* & 595 & 0.19 & -0.81 & -1.35 & 0.15\\
1r69(2) & 1 & -2.90 & -1.23 & -1.72 & 0.16\\
1ris(5) & 12 & -1.75 & -0.57 & -1.65 & 0.18\\ 
1tul(5) & 1 & -3.40 & -0.31 & -1.52 & 0.16\\
1vls(3) & 1 & -2.78 & -0.62 & -1.82 & 0.19\\ 
1who(6) & 1 & -3.56 & -0.21 & -0.79 & 0.19\\
2acy(6) & 1 & -4.75 & -0.25 & -1.74 & 0.15\\
2gdm(5) & 1 & -3.55 & -0.07 & -1.24 & 0.13\\
5pti(3) & 2 & -3.25 & -0.25 & -0.81 & 0.07\\
\hline
Total/average & 17/23 & -3.44 & -0.54 & -1.65 & 0.16 \\
\hline
\end{tabular*}
\end{center}

%\begin{tabular}{|l||l|l||l|l|}
%\hline
% &\multicolumn{2}{l|}{Singular}&\multicolumn{2}{l|}{Plural}\\
%\cline{2-5}
% &English&\textbf{Gaeilge}&English&\textbf{Gaeilge}\\
%\hline\hline
%1st Person&at me&\textbf{agam}&at us&\textbf{againn}\\
%2nd Person&at you&\textbf{agat}&at you&\textbf{agaibh}\\
%3rd Person&at him&\textbf{aige}&at them&\textbf{acu}\\
% &at her&\textbf{aici}& & \\
%\hline
%\end{tabular}

\newpage
{\bf Table G: Comparison with Chebyshev-Expanded hydrophobic potential (CHP) for the decoy sets 4state\_reduced and lmds. The data separated by colons are \textsl{our score : CHP score}}

\begin{center}
%\begin{tabular}{|l|l|l||l|l|l|}
\begin{tabular*}{0.60\textwidth}{@{\extracolsep{\fill}}ccc|ccc}
\multicolumn{3}{c}{4state\_reduced}&\multicolumn{3}{c}{lmds}\\
\hline
target & Rank 1 & $Z$-score &target & Rank 1 & $Z$-score\\
\hline
1ctf & 1:1 & -3.53:-2.9 &1ctf & 1:1 & -3.42:-3.5 \\ 
1r69 & 1:2 & -3.68:-2.4 &1fc2 & 409:9 & 0.91:-2.1 \\
1sn3 & 1:3 & -2.52:-2.1 &1igd & 1:1 & -2.87:-2.8 \\ 
2cro & 2:60 & -3.01:-1.3 &1shf-A & 1:n/a & -2.90:n/a \\
3icb & 1:10 & -2.26:-1.6 &2cro & 1:250 & -3.42:-0.4 \\
4pti & 5:62 & -2.51:-1.7 &2ovo & 16:7 & -1.67:-2.2 \\
4rxn & 4:6 & -2.71:-2.2 &4pti & 6:32 & -2.24:-1.4 \\
\hline
\end{tabular*}
\end{center}

{\bf Table H: Comparison with SOLV and FRST. The data separated by colons are
\textsl{our score : SOLV score : FRST score}}
\begin{center}
\begin{tabular*}{1.00\textwidth}{@{\extracolsep{\fill}}l|lllll}
\hline
decoy set & Rank 1 & $Z$-score &logPB1 & logPB10 & F.E.\\
\hline
4state\_reduced & 4/7:1/7:7/7 & -2.89:-1.7:-4.4 & -1.19:-0.51:-1.62 & -1.88:-1.78:-2.31 & 0.28:0.28:0.43 \\
lattice\_ssfit & 6/8:4/8:8/8 & -4.06:-3.2:-6.7 & -0.41:-0.37:-0.58 & -1.55:-1.20:-1.29 & 0.12:0.10:0.10 \\
lmds & 4/8:2/8:6/8 & -2.75:-1.2:-3.5 & -0.48:-0.49:-0.48 & -1.36:-1.46:-1.44 & 0.14:0.15:0.14 \\ 
\hline
\end{tabular*}
\end{center}

\newpage

{\bf Figure Caption}

Figure 1. A schematic diagram to show the quantities involved in the
  calculation of  of $M_{ijk}$\\

Figure 2. A schematic diagram to show the calculation  of $M_{ijk}$ for equal sized spheres. Fig.2a corresponds to the 
limiting case $M_{ijk}=0.5$ 
whereas Fig.2c presents $M_{ijk}=1$ upper limit. The Fig.2b is intermediate between 
the above mentioned limits, namely $M_{ijk}=0.64$ \\

Figure 3. A plot to compare effective number of neighbours to a residue with its actual buried area obtained by using GETAREA

\end{document}